\documentclass[entropy,article,accept,pdftex,moreauthors]{Definitions/mdpi} 
\usepackage{tensor}
\usepackage{tikz,ifthen}
\usepackage{bbold}
\usepackage{tikz-network}
\usetikzlibrary{patterns,decorations.pathreplacing,calligraphy}
\usepackage{color}
\usepackage{MnSymbol}

\firstpage{1} 
\pubvolume{1}
\issuenum{1}
\articlenumber{0}
\pubyear{2024}
\copyrightyear{2024}
\datereceived{ } 
\daterevised{ } % Comment out if no revised date
\dateaccepted{ } 
\datepublished{ } 
\hreflink{https://doi.org/} % If needed use \linebreak
\newcommand{\titleinfo}{Hilbert space delocalization under random unitary circuits}
\Title{\titleinfo}

\TitleCitation{\titleinfo}

\Author{Xhek Turkeshi $^{1,\dagger}$\orcidA{}, Piotr Sierant $^{2}$\orcidB{}}

\AuthorNames{Xhek Turkeshi, Piotr Sierant}

\AuthorCitation{Turkeshi, X.; Sierant, P.}
\address{%
$^{1}$ \quad turkeshi@thp.uni-koeln.de}

\firstnote{Institut f\"ur Theoretische Physik, Universit\"at zu K\"oln, Z\"ulpicher Strasse 77a, 50937 K\"oln, Germany}  % Current address should not be the same as any items in the Affiliation section.
\secondnote{ICFO-Institut de Ci\`encies Fot\`oniques, The Barcelona Institute of Science and Technology,
Av. Carl Friedrich Gauss 3, 08860 Castelldefels (Barcelona), Spain}

\abstract{
Unitary dynamics of a quantum system initialized in a selected basis state yields, generically, a state that is a superposition of all the basis states. This process, associated with the quantum information scrambling and intimately tied to the resource theory of coherence, may be viewed as a gradual delocalization of the system's state in the Hilbert space. 
This work analyzes the Hilbert space delocalization under dynamics of random quantum circuits, which serve as a minimal model of chaotic dynamics of quantum many-body systems. We employ analytical methods based on the replica trick and Weingarten calculus to investigate the time evolution of the participation entropies which quantify the Hilbert space delocalization. We demonstrate that the participation entropies approach, up to a fixed accuracy, their long-time saturation value in times that scale logarithmically with the system size. Exact numerical simulations and tensor network techniques corroborate our findings. 
}


\definecolor{myyellow}{RGB}{240,188,66}
\definecolor{myorange}{RGB}{255,102,0}
\definecolor{myorangel}{RGB}{255,204,153}
\definecolor{myblue}{RGB}{66,135,245}

\renewcommand{\boxed}[1]{%
  \framebox{\raisebox{0pt}[0.4\baselineskip][0.025\baselineskip]{\hbox to 0.25cm{\hss#1\hss}}}}

\begin{document}

%%%%%%%%%%%%%%%%%%%%%%%%%%%%%%%%%%%%%%%%%%
\section{Introduction}
Starting from a vector of a generically chosen basis of the Hilbert space, unitary quantum dynamics generate a superposition spanning the entire basis of the Hilbert space. This phenomenon, referred to as Hilbert space delocalization, can be viewed as the spreading of the many-body wave function over the Hilbert space under quantum dynamics. Hilbert space delocalization is tied to non-equilibrium processes in quantum mechanics and, hence, pivotal in our understanding of quantum foundations~\cite{Zu2003}, quantum technologies~\cite{nielsen00,preskill_quantum_2018}, and condensed matter theory~\cite{amico2008entanglementinmanybody,La2016,DAlessio2016}. 
Being intimately connected to the resource theories of quantum coherence and entanglement~\cite{ChGo2019,jamir}, Hilbert space delocalization provides a valuable tool for the characterization of quantum phases of matter~\cite{stephan2009,stephan2010,alcaraz2013universal,luitz2014improvingentanglementand,LaRa2014,luitz2014shannonrenyientropy}, quantum chaos and thermalization~\cite{srednicki,deutsch,pappalardi,fava2023designs,pappalardi2023general,pappalardi2024,backer,pausch2021chaosanderdoficity} and their violations in non-ergodic systems -- including monitored systems~\cite{Fisher2023,potter2022quantumsciencesandtechnology,lunt2022quantumsciencesandtechnology,Sierant2022,Sierant2022_2,chahine2023entanglement} and disorder-induced localization~\cite{abanin,sierant2024manybody,Evers2000,Evers2008,Rodriguez2009,Rodriguez2010,DeLuca2013,mace2019multifractal,Pietracaprina2021,altshuler2024renormalization,scoquart2024role,vanoni2023renormalization,sierant2023universality,vanoni2024analysis,detomasi}. 

Natural quantifiers for localization and delocalization properties of a state $\rho =|\Psi\rangle\langle \Psi|$ over the many-body Hilbert space are the inverse participation ratios (IPR) and the participation entropies, that measure the spreading of the state distribution $p_n \equiv \langle n|\rho|n\rangle$ over the basis $\mathcal{B}=\{|n\rangle\}$. 
The IPR and participation entropy are given respectively, by~\footnote{We note the connection with the relative entropies of coherence~\cite{Baumgratz} $C_q(\rho,\mathcal{B}) \equiv S_q - (1-q)^{-1}\ln\mathrm{tr}(\rho^q)$, a scalable quantifier in the quantum coherence resource theory. In particular, for pure states, these two quantities coincide $C_q(\rho,\mathcal{B}) = S_q$.} 
\begin{equation}
I_q \equiv \sum_{n\in \mathcal{B}} \left(\langle n| \rho| n \rangle\right)^q = \sum_{n\in \mathcal{B}} p_n^q \;,\qquad S_q = \frac{1}{1-q} \ln \left[I_q\right]\;.
\label{eq:def}
\end{equation}
Despite the explicit basis dependence, extensive studies demonstrated that the IPR captures structural properties of quantum matter, including universal behavior at phase transitions~\cite{luitz2014universalbehaviorbeyond,Sierant2022}. 
However, the investigations of IPR and participation entropies have focused so far on the equilibrium and stationary state features, leaving the question of the time evolution of quantum coherence and, hence, of Hilbert space delocalization unresolved.

In this work, we investigate Hilbert space delocalization under the dynamics of $(1+1)D$ quantum circuits comprising 2-qudit Haar random gates arranged in a brick-wall pattern~\cite{Fisher2023}.
The locality and unitarity of this setup constitute minimal requirements for chaotic evolution in many-body systems~\cite{nahum2017quantum,nahum2018operator,andrea0,andrea00,andrea1,zhou2019,zhou2020,lorenzo1,keyserlingks1,keyserlingks,kheman,Snderhauf2019,piroli2020,bruno2,pavel,bruno0,peter1,entanglementgrowth,rampp2023haydenpreskill,boundary,bulchandani2023randommatrix,deluca2023universality}, hence allow us for a phenomenological understanding of the Hilbert space delocalization 
under generic quantum dynamics.
Our analysis combines rigorous analytical methods, based on the mapping between average of IPR over the random circuits and a statistical mechanics model, together with exact numerical simulations, including tensor network techniques~\cite{Schollw_ck_2011,Ors2019,Or_s_2014,Bauls2023}~\footnote{The implementation, based on the open-source library~\textsc{ITensor}~\cite{itensor,itensor-r0.3}, and available at [Reference available for published version].}. 
We find 
that the saturation of participation entropy $S_q$ to the long-time stationary value, equal to the random Haar state average $\widetilde{S}^{\mathrm{H}}_q$, occurs exponentially quickly in time (i.e. in circuit depth)
\begin{equation}
S_q = \widetilde{S}^{\mathrm{H}}_q - \alpha_q N (2K_d)^{t-1},
\label{eq:main}
\end{equation}
where $K_d>0$ is a constant characterizing the properties of 2-qudit gates and $\alpha_q>0$ is a constant. We find an exact expression for the constant $K_d$ for Haar random 2-qudit gates, and provide analytic arguments for the value of $\alpha_q$ in the limit of large on-site Hilbert space dimension. In other words, Eq.~\eqref{eq:main} implies that the participation entropy $S_q$ approaches its stationary value, up to a fixed accuracy, at time scaling logarithmically with the system size, $\tau_\mathrm{HSD}\sim \ln(N)$. 

The paper is structured as follows. In Sec.~\ref{sec:randtot}, we outline the toolbox employed in the rest of the paper, while in Sec.~\ref{sec:radomHaar}, we discuss the stationary value of participation entropy that the deep quantum circuit reaches.  Sec.~\ref{sec:randcirc} presents the core of our analytical approach based on the mapping of IPR calculation to a statistical mechanics problem. We resolve the R\'enyi-2 participation entropy evolution in Sec.~\ref{sec:randcircA}, where we present the analytic prediction.
These findings are further corroborated by the numerical analysis in Sec.~\ref{sec:randcircC} for generic R\'enyi index $q$. 
We present an outlook of our work and a discussion of its further implications in Sec.~\ref{sec:conclusion}.
The more technical parts of the paper are detailed in the Appendix.  In App.~\ref{app:A} we present a self-contained discussion about Haar averages, in App.~\ref{app:B} we review the interface problem related to the entanglement propagation, and in App.~\ref{app:C} briefly discuss the numerical implementations.

\section{Methods}
\label{sec:randtot}
We begin presenting the analytical toolbox we will employ in the rest of the paper. The latter is based on the replica trick, the superoperator formalism~\cite{breuer2002theory}, and the Weingarten calculus~\cite{Weingarten1978,Collins2006,Kostenberger2021,Collins2021}, see also Ref.~\cite{zhou2019,piroli2020}. 
Since $\langle n|\bullet|m\rangle = \mathrm{tr}(|m\rangle\langle n|(\bullet))$ and $\mathrm{tr}(A B)^q = \mathrm{tr}(A^{\otimes q}B^{\otimes q})$, the IPR definition \eqref{eq:def} implies that $I_q = \mathrm{tr}(\Lambda_\mathcal{B} \rho^{\otimes q})$, where $\Lambda_{\mathcal{B}} = \sum_{n \in \mathcal{B}} |n\rangle\langle n|^{\otimes q}$ is a replica operator acting on $\mathcal{H}_N^{\otimes q}$. 
Throughout this manuscript we will study qudits with local Hilbert space dimension $d$, hence  $\mathcal{H}_N\simeq \mathbb{C}^{d^N}$. For concreteness, we will fix $\mathcal{B}$ to be the computational basis, where $|n\rangle = |b^1,\dots,b^N\rangle$ with $b_j=0,\dots,d-1$ for each $j$. %
Nevertheless, due to the invariance of the considered circuits under local basis rotations, our results remain valid for any basis of $\mathcal{H}_N$ obtained by a unitary transformation $U=U_1 \otimes \cdots \otimes U_N$ of the computational basis, where $U_k$ belongs to the unitary group $\mathcal{U}(d)$ for a single qudit.
For the computational basis, $\Lambda_{\mathcal{B}} = \bigotimes_{j=1}^N\Lambda_q^{(j)} $ which defines the ``book'' boundary condition for each qudit~\cite{turkeshi2023measuring,turkeshi2024}
\begin{equation}
    \Lambda_q^{(j)} \equiv  \sum_{b^j=0,\dots,d-1} (|b^j\rangle\langle b^j|)^{\otimes q}\;,
\end{equation}
where the superscript in $\Lambda_q^{(j)}$ bookkeeps the 
number of the affected qudit.
For later convenience, we recast the problem in the superoperator formalism. For any operator $U$ and $A$ we have $A = \sum_{n,m=0}^{d^N-1} A_{n,m} |n\rangle\langle m|\mapsto |A\rrangle = \sum_{n,m=0}^{d^N-1} A_{n,m} |n,m\rrangle$, and $U A U^\dagger \mapsto (U\otimes U^*) |A\rrangle $~\footnote{We denote with $(\bullet)^*$ and $(\bullet)^\dagger$, respectively, the complex and hermitian conjugation of $\bullet$.}. In this representation %$\langle B|A\rangle = \mathrm{tr}(B^\dagger A)$, 
$\llangle B|A\rrangle = \mathrm{tr}(B^\dagger A)$, 
so the IPR can be written as 
\begin{equation}
    I_q = \llangle \Lambda_q^{(1)},\Lambda_q^{(2)},\dots, \Lambda_q^{(N)} | \rho^{\otimes q}\rrangle. 
\end{equation}
For convenience, we introduce a graphical notation. We implicitly define it for the replica boundary $\Lambda_q$ and the replica state $\rho^{\otimes q}$ via the IPR 
\begin{equation}
    I_q \equiv  \llangle \Lambda_q^{(1)},\Lambda_q^{(2)},\dots, \Lambda_q^{(N)} |\rho^{\otimes q}\rrangle\; = 
    \begin{tikzpicture}[baseline=(current  bounding  box.center), scale=0.4]
    \foreach \i in {1,...,6} {
      \draw[ultra thick] (-4.5+1.5*\i,1.5) -- (-4.5+1.5*\i,0);
      \draw[fill=white,very thick](-4.5+1.5*\i,2.0) circle (0.66) node {$\Lambda_q$};
    }
    \draw[ultra thick, fill=white, rounded corners=3pt] (-4.0,-1) rectangle node{$\rho^{\otimes q}$}(5.5,1);
    \end{tikzpicture}\;.
\end{equation}
Lastly, since we will consider the unitary evolution $|\rho\rrangle = (U \otimes U^*) |\rho_0\rrangle$ for some $|\rho_0\rrangle$ selected as the initial state and unitary operation $U$ acting on the Hilbert space $\mathcal{H}_N$, we introduce the graphical notation for $U$ and $U^*$, respectively, 
\begin{equation}
    [U]_{i_1,\dots,i_N}^{j_1,\dots,j_N} = \begin{tikzpicture}[baseline=(current  bounding  box.center), scale=0.4]
    \foreach \i in {1,...,6} {
    \pgfmathsetmacro{\col}{ifthenelse(\i< 3,"$i_\i$",ifthenelse(\i==6,"$i_N$","\dots"))}
    \pgfmathsetmacro{\row}{ifthenelse(\i< 3,"$j_\i$",ifthenelse(\i==6,"$j_N$","\dots"))}
      \draw[ thick]  (-4.5+1.5*\i,1.5) node[left]{\row}-- (-4.5+1.5*\i,-1.5) node[left]{\col};
    }
    \draw[ thick, fill=myorange, rounded corners=3pt] (-3.5,-0.9) rectangle (5.,0.9);
    \end{tikzpicture}\;,\qquad [U^*]_{i_1,\dots,i_N}^{j_1,\dots,j_N} = \begin{tikzpicture}[baseline=(current  bounding  box.center), scale=0.4]
    \foreach \i in {1,...,6} {
    \pgfmathsetmacro{\col}{ifthenelse(\i< 3,"$i_\i$",ifthenelse(\i==6,"$i_N$","\dots"))}
    \pgfmathsetmacro{\row}{ifthenelse(\i< 3,"$j_\i$",ifthenelse(\i==6,"$j_N$","\dots"))}
      \draw[ thick]  (-4.5+1.5*\i,1.5) node[left]{\row}-- (-4.5+1.5*\i,-1.5) node[left]{\col};
    }
    \draw[ thick, fill=myyellow, rounded corners=3pt] (-3.5,-0.9) rectangle (5.,0.9);
    \end{tikzpicture}\;.
\end{equation}
We will omit the legs 
indices, which can be inferred from the context. Moreover, we will typically reserve thick (thin) lines for multi- (single)-replica objects.

\section{Delocalization properties of random Haar states}
\label{sec:radomHaar}
We begin our discussion reviewing the Hilbert space delocalization of random states~\cite{backer,turkeshi2023measuring} 
which correspond to the stationary ensemble of states obtained under the action of sufficiently deep random circuits. 
Uniformly distributed random states in the Hilbert space $|\rho_\mathrm{Haar}\rrangle = (U\otimes U^*)|\rho_0\rrangle$ are obtained from a reference state $|\rho_0\rrangle =|\Psi_0,\Psi_0\rangle$ acting with a Haar distributed unitary $U\in \mathcal{U}(d^N)$, where $d$ is the qudit local Hilbert space dimension and $N$ is the total number of qudits. 
Let us compute the average IPR over the Haar ensemble, which reads, using the graphical notation introduced in Sec.~\ref{sec:randtot},
\begin{equation}
    \overline{I}^{\mathrm{H}}_q \equiv \mathbb{E}_\mathrm{Haar} \left[\llangle \Lambda_q^{(1)},\Lambda_q^{(2)},\dots, \Lambda_q^{(N)} | (U\otimes U^*)^{\otimes q}|\rho_0^{\otimes q}\rrangle\right]\; = 
    \begin{tikzpicture}[baseline=(current  bounding  box.center), scale=0.4]
    \foreach \i in {1,...,6} {
      \draw[ultra thick] (-4.5+1.5*\i,1.5) -- (-4.5+1.5*\i,-1.5);
      \draw[fill=white,very thick](-4.5+1.5*\i,2.0) circle (0.66) node {$\Lambda_q$};
    }
    \draw[ultra thick, fill=myblue, rounded corners=3pt] (-4.0,-1) rectangle (5.5,1);
    \draw[ultra thick, fill=white, rounded corners=3pt] (-4.0,-1.5) rectangle node{$\rho_0^{\otimes q}$}(+5.5,-3.5);
    \end{tikzpicture}\;,
\end{equation}
where $\mathbb{E}_\mathrm{Haar}$ denotes the expectation value over the unitary group $\mathcal{U}(d^N)$ taken with the Haar measure.
Here, we have defined, via the linearity of the average and of the expectation value, the $q$-replica transfer matrix on $N$ qudits
\begin{equation}
W^{(q)}_N\equiv\mathbb{E}_\mathrm{Haar}\left[(U\otimes U^*)^{\otimes q}\right] = 
    \mathbb{E}_\mathrm{Haar}\left[\begin{tikzpicture}[baseline=(current  bounding  box.center), scale=0.4]
    \def\k{0.0};
    \draw [decorate,decoration={brace}] (-3,+1.5) -- node[above]{$q$}(-1,3.5);
    \foreach \k in {8,7,...,1} {
    \pgfmathsetmacro{\col}{ifthenelse(int(mod(\k,2))==1,"myorange","myyellow"}
    \foreach \i in {1,...,6} {
      \draw[ thick] (-3.5+\i+0.2*\k,1.5+0.2*\k) -- (-3.5+\i+0.2*\k,-1.5+0.2*\k);
    }
    \draw[ thick, fill=\col, rounded corners=3pt] (-3+0.2*\k,-1+0.25*\k) rectangle (3+0.2*\k,1+0.2*\k);
    }
    \end{tikzpicture}\right]\, \equiv  \begin{tikzpicture}[baseline=(current  bounding  box.center), scale=0.4]
    \foreach \i in {1,...,6} {
      \draw[ultra thick] (-3.5+\i,1.5) -- (-3.5+\i,-1.5);
    }
    \draw[ultra thick, fill=myblue, rounded corners=3pt] (-3,-1) rectangle (3,1);
    \end{tikzpicture} \;. 
    \label{eq:Wdef}
\end{equation}
The explicit formula for $W^{(q)}_N$ can be obtained by a direct evaluation of the Haar average, as detailed in App.~\ref{app:A}. 
The final result is expressed in terms of the permutation operators $|\sigma\rrangle_j$ acting on each qudit $j$, with the matrix elements in the superoperator computational basis given as
\begin{equation}
    \llangle b_1^j,\bar{b}_1^j,b_2^j,\bar{b}_2^j,\dots,b_q^j,\bar{b}_q^j|\sigma\rrangle_j = \prod_{k=1}^q \delta_{b_k^j,\bar{b}_k^{\sigma(j)}}\;,
\end{equation}
fixed by the permutation $\sigma\in\mathcal{S}_q$. Employing the Weingarten function $\mathrm{Wg}(D;\sigma)$~\footnote{
Denoting the projectors of the irreducible representations of the permutation group $\mathcal{S}_k$ onto the $D$ dimensional Hilbert space by $\Pi_\lambda$, their dimensions by $d_\lambda$ and their character functions by $\chi_\lambda$, the Weingarten function can be written as sums over the integer partitions of $k$
\begin{equation}
    \mathrm{Wg}(D;\sigma) = \sum_{\lambda \vdash k} \frac{
    d_\lambda^2}{(k!)^2} \frac{\chi_\lambda(\sigma)}{\mathrm{tr}(\Pi_\lambda)}\;,
\end{equation}
see Ref.~\cite{Weingarten1978,Collins2021} fur further details.} allows us to define the dual states  $| \tilde{\sigma}\rrangle_{1,2,\dots,N} = \sum_{\tau\in \mathcal{S}_q}  \mathrm{Wg}(d^N; \sigma \tau^{-1}) |\tau \rrangle_1\otimes \cdots \otimes | \tau \rrangle_N$ acting on the whole system of $N$ qudits. With these states, the transfer matrix reads
\begin{equation}
\label{eq:WWW}
    W^{(q)}_N =\sum_{\sigma\in\mathcal{S}_q}  | \tilde{\sigma}\rrangle_{1,2,\dots,N} \llangle \sigma|_1\otimes \cdots \otimes \llangle \sigma|_N\equiv \sum_{\sigma\in\mathcal{S}_q}\begin{tikzpicture}[baseline=(current  bounding  box.center), scale=0.4]
\def\eps{0.5};
\foreach \i in {1,...,6} {
  \draw[ultra thick] (-3.+.9*\i,1.5) -- (-3.+.9*\i,-0.75);
  \draw[very thick](-3+.9*\i,-1.5) -- (-3+.9*\i,-2.5);
  \draw[fill=myorangel,very thick](-3+.9*\i,-1.5) circle (0.4) node {$\sigma $};
}
\draw[ultra thick, fill=myorangel, rounded corners=3pt] (-3,+1) rectangle (3,-0.75) (0,0.25) node {$\tilde{\sigma}$};
\end{tikzpicture}\,.
\end{equation}
We are now in position to compute the IPR. Since the initial state $\rho_0$ is pure, $\llangle \sigma,\dots,\sigma|\rho_0^{\otimes q}\rrangle = 1$. Furthermore, using that $\sum_{\sigma} \mathrm{Wg}(d^N;\sigma \tau^{-1}) = (d^N-1)!/(d^N+q-1)!$~\cite{Kostenberger2021} we have
\begin{equation}
\begin{split}
    \overline{I_q} &= \sum_{\sigma\in\mathcal{S}_q}\begin{tikzpicture}[baseline=(current  bounding  box.center), scale=0.4]
\def\eps{0.5};
\foreach \i in {1,...,6} {
      \draw[ultra thick] (-4.5+1.5*\i,1.5) -- (-4.5+1.5*\i,0);
      \draw[fill=white,very thick](-4.5+1.5*\i,2.2) circle (0.66) node {$\Lambda_q$};
      \draw[ultra thick] (-4.5+1.5*\i,-1.75) -- (-4.5+1.5*\i,-2.5);
      \draw[fill=myorangel,very thick](-4.5+1.5*\i,-1.75) circle (0.5) node {$\sigma$};
    }
    \draw[ultra thick, fill=myorangel, rounded corners=3pt] (-4.0,-1) rectangle (5.5,1) (0.25,0.0) node {$\tilde{\sigma}$};
    \draw[ultra thick, fill=white, rounded corners=3pt] (-4.0,-2.5) rectangle node{$\rho_0^{\otimes q}$}(+5.5,-4);
\end{tikzpicture}\, = \sum_{\sigma\in\mathcal{S}_q} \begin{tikzpicture}[baseline=(current  bounding  box.center), scale=0.4]
\def\eps{0.5};
\foreach \i in {1,...,6} {
      \draw[ultra thick] (-4.5+1.5*\i,1.5) -- (-4.5+1.5*\i,0);
      \draw[fill=white,very thick](-4.5+1.5*\i,2.2) circle (0.66) node {$\Lambda_q$};
    }
    \draw[ultra thick, fill=myorangel, rounded corners=3pt] (-4.0,-1) rectangle (5.5,1) (0.25,0.0) node {$\tilde{\sigma}$};
\end{tikzpicture}\\ &= \frac{1}{d^N(d^N+1)\cdots (d^N+q-1)} \sum_{\tau \in \mathcal{S}_q} \left( 
\begin{tikzpicture}[baseline=(current  bounding  box.center), scale=0.4] 
\draw[ultra thick] (0,1.2) -- (0,-0.5);
\draw[fill=white,very thick](0,1.2) circle (0.66) node {$\Lambda_q$};
\draw[fill=myorangel,very thick](0,-0.75) circle (0.5) node {$\tau$};
\end{tikzpicture}\right)^N = \frac{q!}{(d^N+1)\cdots (d^N+q-1)}\;,
\end{split}
\label{eq:IPRann}
\end{equation}
where we used $|\mathcal{S}_q|=q!$ and $\llangle \Lambda_q |\tau\rrangle = d$ for all $\tau\in \mathcal{S}_q$. 
From this calculation, it follows that the \emph{annealed averaged} R\'enyi entropy for the Haar states is given by 
\begin{equation}
    \tilde{S}^{\mathrm{H}}_q\equiv  \frac{1}{1-q}\ln\left[\overline{I}^{\mathrm H}_q\right] = \frac{\ln[(d^N+1)\cdots (d^N+q-1)] - \ln[q!] }{q-1}
    \stackrel{ N \to \infty}{\longrightarrow}
    N \ln[d]  - \frac{1}{q-1}\ln[q!]  \;. 
    \label{eq:PEann}
\end{equation}
As expected and already discussed in~\cite{backer}, the Haar random states are (almost) fully delocalized over the many-body basis $\mathcal B$. Indeed, \eqref{eq:PEann} shows that $\tilde{S}^{\mathrm{H}}_q$ differs only by a sub-leading constant term $(q-1)^{-1}\ln[q!]$ from the maximal value of participation entropy $N \ln(d)$ achieved for a unifromly distributed state with $\langle n| \rho| n \rangle = d^{-N}$.
Since the logarithm is non-linear, in principle one should expect corrections when considering the \emph{quenched average} of the participation entropy  $\overline{S}^{\mathrm H}_q \equiv \mathbb{E}_\mathrm{Haar}[S_q]$. We briefly estimate these fluctuations via the variance~\footnote{For the explicit expression for $\overline{I_q^r}$, see Ref.~\cite{turkeshi2023measuring}.} 
\begin{equation}
    \mathrm{std}(I_q)\equiv \sqrt{\overline{I_q^2}- \left(\overline{I_q}\right)^2}\;.
\end{equation}
For this expression, we need to compute $\overline{I_q^2}$, which requires the use of $2q$ replicas of $\mathcal{H}_N$, and the boundary conditions $\llangle \Lambda_q\otimes \Lambda_q|$ acting on each site. 
Algebraic manipulations analogous to the ones described above give
\begin{equation}
    \overline{I_q^2} = \frac{ (d^N-1)!}{(d^N+2q-1)!} \sum_{\tau \in \mathcal{S}_{2q}} \left( 
\begin{tikzpicture}[baseline=(current  bounding  box.center), scale=0.4] 
\draw[ultra thick] (0,1.2) -- (0,-0.5);
\draw[fill=white,very thick](0,1.2) ellipse (2 and 1) node {$\Lambda_q\otimes \Lambda_q $};
\draw[fill=myorangel,very thick](0,-0.75) circle (0.5) node {$\tau$};
\end{tikzpicture}\right)^N = \frac{(q!)^2 (d^N-1) + (2q)!}{(d^N+1)\cdots (d^N+2 q-1)}\;.
\end{equation}
It follows that $\mathrm{std}(I_q)\sim O\left(d^{(1-2q)N/2}\right)$, and in the thermodynamic limit, these fluctuations are irrelevant. In particular, in the scaling limit  $ N \to \infty$, the annealed and the quenched averages coincide, $ \tilde{S}^{\mathrm H}_q - \overline{S}^{\mathrm H}_q 
 \stackrel{ N \to \infty}{\longrightarrow} 0 $, and with probability equal to unity, approximate the value of participation entropy calculated for a single Haar random state with an arbitrarily small fixed accuracy. 

\section{Hilbert space delocalization in brick-wall quantum circuits}
\label{sec:randcirc}
After this preliminary discussion, relevant to the deep circuit limit, we now discuss the Hilbert space delocalization under random quantum circuits. 
As the initial state we fix $|\Psi_0\rangle = |0\rangle^{\otimes N}$, and study the approach of the annealed and quench average participation entropy to the asymptotic value $\tilde{S}_q^\mathrm{H}= (1-q)^{-1}\ln[I_q^\mathrm{H}] $, with $I_q^\mathrm{H} \equiv q! (d^N)!/ (d^N+q-1)!$, cf.~Eqs.~\eqref{eq:IPRann} and~\eqref{eq:PEann}.
The evolution operator corresponding to the brick-wall quantum circuit of depth $t$ (referred to also as time) 
is given by 
\begin{equation}
    U_t = \prod_{\tau=0}^{t-1} U^{(\tau)}\;,\qquad U^{(\tau)} = \prod_{i=0}^{N/2-\lfloor{\tau/2}\rfloor} U_{2i+\lfloor{\tau/2}\rfloor,2i+1+\lfloor{\tau/2}\rfloor}\;,
\end{equation}
where $\lfloor{\bullet}\rfloor$ is the integer part of $\bullet$. Here, each $U_{i,j}$ is an independent identically distributed (i.i.d.) Haar random unitary from $\mathcal{U}(d^2)$ acting on the $i$-th and $j$-th qudit. 
Our goal is to compute the 
\begin{equation}
    \overline{I}_q\equiv \mathbb{E}_\mathrm{Haar}\left[\llangle \Lambda_q,\dots,\Lambda_q| (U_t\otimes U^*_t)^{\otimes q} | \psi_0,\psi_0,\dots,\psi_0\rrangle\right]\;,
\end{equation}
where the $N$ copies of $|\psi_0\rrangle = |0\rangle^{\otimes 2q}$ implement the initial condition $|\Psi_0\rangle$ on the replica space. We note that the $\mathbb{E}_\mathrm{Haar}$ denotes here the average over the realizations of the two-body gates $U_{i,j}$ which comprise the considered brick-wall circuit.
We adapt the graphical notation of Sec.~\ref{sec:randtot} to the two-body gates
\begin{equation}
U_{i,j}^{k,l}=
\begin{tikzpicture}[baseline=(current  bounding  box.center), scale=1]
\def\eps{0.5};
\draw[thick] (-1.75,-0.5)node[left]{$i$} -- (-1.75,0.5)node[left]{$k$};
\draw[thick] (-1.25,-0.5)node[right]{$j$} -- (-1.25,0.5)node[right]{$l$};
\draw[thick, fill=myorange, rounded corners=2pt] (-1.9,0.25) rectangle (-1.1,-0.25);
\end{tikzpicture}\,,
\qquad
\left(U^{*}\right)_{i,j}^{k,l}=
\begin{tikzpicture}[baseline=(current  bounding  box.center), scale=1]
\def\eps{0.5};
\draw[thick] (-1.75,-0.5)node[left]{$i$} -- (-1.75,0.5)node[left]{$k$};
\draw[thick] (-1.25,-0.5)node[right]{$j$} -- (-1.25,0.5)node[right]{$l$};
\draw[thick, fill=myyellow, rounded corners=2pt] (-1.9,0.25) rectangle (-1.1,-0.25);
\end{tikzpicture}\;,\qquad W^{(q)}_2 = 
\begin{tikzpicture}[baseline=(current  bounding  box.center), scale=1]
\def\eps{0.5};
\draw[ultra thick] (-1.75,-0.5) -- (-1.75,0.5);
\draw[ultra thick] (-1.25,-0.5)-- (-1.25,0.5);
\draw[ultra thick, fill=myblue, rounded corners=2pt] (-1.9,0.25) rectangle (-1.1,-0.25);
\end{tikzpicture}\,= \sum_{\sigma \in \mathcal{S}_q}
\begin{tikzpicture}[baseline=(current  bounding  box.center), scale=1]
\def\eps{0.5};
\draw[very thick] (-1.75,0.5)node[left]{} -- (-1.75,0.75)node[left]{};
\draw[very thick] (-1.25,0.5)node[left]{} -- (-1.25,0.75)node[left]{};
\draw[very thick, fill=myorangel, rounded corners=2pt] (-2,0.5) rectangle (-1.0,0) (-1.5,0.25) node {$\tilde{\sigma}$};
\draw[fill=myorangel,ultra thick](-1.75,-0.3) circle (0.2) node {$\sigma $};
    \draw[very thick](-1.75,-0.5) -- (-1.75,-0.75);
\draw[fill=myorangel,ultra thick](-1.25,-0.3) circle (0.2) node {$\sigma $};
    \draw[very thick](-1.25,-0.5) -- (-1.25,-0.75);
\end{tikzpicture}\,,
\label{eq:W2}
\end{equation}
and recall that 
the two-body gates $U_{i,j}$ are i.i.d. variables. It follows that their averages factorize and 
the transfer matrix corresponding to the circuit of depth $t$ reads 
\begin{equation}
    \mathcal{W}_t\equiv \mathbb{E}_\mathrm{Haar}\left[\begin{tikzpicture}[baseline=(current  bounding  box.center), scale=0.4]
    \def\k{0.0};
    \def\eps{0};
    \def\shift{0};
    \draw [decorate,decoration={brace}] (-14,+.5) -- node[above]{$q$}(-12,2.5);
    \draw [decorate,decoration={brace}] (-14,-9) -- node[left]{$t$}(-14,+.3);
    \foreach \kk[evaluate=\kk as \k using 0.25*\kk] in {5,4,3,2,1,0} {
  \pgfmathsetmacro{\col}{ifthenelse(int(mod(\kk,2))==0,"myorange","myyellow"}
  \foreach \i in {1,...,3}{
    \draw[thick] (-1.-4*\i+\k+\shift,1+\k) -- (-1.-4*\i+\k+\shift,-9+\k);
    \draw[thick] (+1.-4*\i+\k+\shift,1+\k) -- (+1.-4*\i+\k+\shift,-9+\k);
  }
  \foreach \jj[evaluate=\jj as \j using -(ceil(\jj/2)-\jj/2), evaluate=\j as \fin using 3+\j] in {0,...,4} {
    \foreach \i in {1,...,\fin}{        
      \draw[thick, fill=\col, rounded corners=1pt] (-1.4-4*\i+4*\j+\k+\shift,+0.5+\eps-2*\jj+\k) rectangle (1.4-4*\i+4*\j+\k+\shift,-0.5+\eps-2*\jj+\k);
    }
  }
} \end{tikzpicture}\right] =
\begin{tikzpicture}[baseline=(current  bounding  box.center), scale=0.4]
    \def\k{0.0};
    \def\eps{0};
    \def\shift{0};
    \draw [decorate,decoration={brace},thick] (-14,-9) -- node[left]{$t$}(-14,+.3);
    \foreach \kk[evaluate=\kk as \k using 0.25*\kk] in {0} {
  \pgfmathsetmacro{\col}{ifthenelse(int(mod(\kk,2))==0,"myblue","myblue"}
  \foreach \i in {1,...,3}{
    \draw[ultra thick] (-1.-4*\i+\k+\shift,1+\k) -- (-1.-4*\i+\k+\shift,-9+\k);
    \draw[ultra thick] (+1.-4*\i+\k+\shift,1+\k) -- (+1.-4*\i+\k+\shift,-9+\k);
  }
  \foreach \jj[evaluate=\jj as \j using -(ceil(\jj/2)-\jj/2), evaluate=\j as \fin using 3+\j] in {0,...,4} {
    \foreach \i in {1,...,\fin}{        
      \draw[ultra thick, fill=\col, rounded corners=1pt] (-1.4-4*\i+4*\j+\k+\shift,+0.5+\eps-2*\jj+\k) rectangle (1.4-4*\i+4*\j+\k+\shift,-0.5+\eps-2*\jj+\k);
    }
  }
} \end{tikzpicture}\;.
\end{equation}
Contracting with the replica boundaries, and recalling the decomposition for $W_2^{(q)}$ in Eq.~\eqref{eq:W2}, we get the final expression
\begin{equation}
    \overline{I_q}(t) =
    \begin{tikzpicture}[baseline=(current  bounding  box.center), scale=0.4]
    \def\k{0.0};
    \def\eps{0};
    \def\shift{0};
      \foreach \i in {1,...,6} {
    \draw[fill=white,very thick](-1-2*\i+\k+\shift,1.75+\k) circle (0.75) node {$\Lambda_q $};
    \draw[fill=white,very thick](-1-2*\i+\k+\shift,-9.75+\k) circle (0.75) node {$\psi_0 $};
  }
    \foreach \kk[evaluate=\kk as \k using 0.25*\kk] in {0} {
  \pgfmathsetmacro{\col}{ifthenelse(int(mod(\kk,2))==0,"myblue","myblue"}
  \foreach \i in {1,...,3}{
    \draw[ultra thick] (-1.-4*\i+\k+\shift,1+\k) -- (-1.-4*\i+\k+\shift,-9+\k);
    \draw[ultra thick] (+1.-4*\i+\k+\shift,1+\k) -- (+1.-4*\i+\k+\shift,-9+\k);
  }
  \foreach \jj[evaluate=\jj as \j using -(ceil(\jj/2)-\jj/2), evaluate=\j as \fin using 3+\j] in {0,...,4} {
    \foreach \i in {1,...,\fin}{        
      \draw[ultra thick, fill=\col, rounded corners=1pt] (-1.4-4*\i+4*\j+\k+\shift,+0.5+\eps-2*\jj+\k) rectangle (1.4-4*\i+4*\j+\k+\shift,-0.5+\eps-2*\jj+\k);
    }
  }
} \end{tikzpicture}\; =
\left[\frac{(d^2)!}{(d^2+q-1)!}\right]^{N/2}\begin{tikzpicture}[baseline=(current  bounding  box.center), scale=0.4]
    \def\k{0.0};
    \def\eps{0};
    \def\shift{0};
      \foreach \i in {1,...,3} {
    \draw[fill=white,very thick](-1.4-4*\i+\k+\shift,+0.0+\eps-\k) rectangle (1.4-4*\i+\shift,-1+\eps+\k) 
        (0.0-4*\i-\k+\shift,-0.5) node {$+$};
  }
    \foreach \kk[evaluate=\kk as \k using 0.25*\kk] in {0} {
  \pgfmathsetmacro{\col}{ifthenelse(int(mod(\kk,2))==0,"myblue","myblue"}
  \foreach \i in {1,...,3}{
    \draw[ultra thick] (-1.-4*\i+\k+\shift,1+\k-2) -- (-1.-4*\i+\k+\shift,-8.+\k);
    \draw[ultra thick] (+1.-4*\i+\k+\shift,1+\k-2) -- (+1.-4*\i+\k+\shift,-8+\k);
  }
  \foreach \jj[evaluate=\jj as \j using -(ceil(\jj/2)-\jj/2), evaluate=\j as \fin using 3+\j] in {4} {
    \foreach \i in {1,...,\fin}{        
      \draw[ultra thick, fill=myorangel, rounded corners=1pt] (-1.4-4*\i+4*\j+\k+\shift,+0.5+\eps-2*\jj+\k) rectangle (1.4-4*\i+4*\j+\k+\shift,-0.5+\eps-2*\jj+\k) (1.4-1.33-4*\i+4*\j+\k+\shift,-0.5+\eps-2*\jj+\k+0.5)node{$\tilde{+}$}  ;
    }
  }
  \foreach \jj[evaluate=\jj as \j using -(ceil(\jj/2)-\jj/2), evaluate=\j as \fin using 3+\j] in {1,...,3} {
    \foreach \i in {1,...,\fin}{        
      \draw[ultra thick, fill=\col, rounded corners=1pt] (-1.4-4*\i+4*\j+\k+\shift,+0.5+\eps-2*\jj+\k) rectangle (1.4-4*\i+4*\j+\k+\shift,-0.5+\eps-2*\jj+\k);
    }
  }
} \end{tikzpicture}\;,
\label{eq:Wtensor}
\end{equation}
where in the last step we used that $\llangle \sigma | \psi_0\rrangle = 1$ for each site, $(\llangle \Lambda_q|_i \otimes \llangle \Lambda_q|_{i+1}) |\tilde{\sigma} \rrangle_{i,i+1} = d^2$, and we defined $|+\rrangle_{i,i+1} = \sum_{\sigma \in \mathcal{S}_q} |\sigma\rrangle_i\otimes |\sigma\rrangle_{i+1}$ and $|\tilde{+}\rrangle_{i,i+1} =\sum_{\sigma \in \mathcal{S}_q} |\tilde{\sigma}\rrangle_{i,i+1}$.
Expression Eq.~\eqref{eq:Wtensor} is one of the main results of this work, as it already allows for efficient tensor network implementations~\cite{itensor}.
At the same time, Eq.~\eqref{eq:Wtensor} can be viewed as the partition function of a statistical mechanics model once the $W_2^{(q)}$ are replaced with their decomposition Eq.~\eqref{eq:W2}.
To highlight this interpretation, we can unravel the upper boundary condition as 
\begin{equation}
    |\mathfrak{Q}_\mathrm{in}\rrangle = \sum_{\{\sigma_j\in\mathcal{S}_q\}}|\mathfrak{Q}_\mathrm{in}(\{\sigma_j\})\rrangle  \;\qquad |\mathfrak{Q}_\mathrm{in}(\{\sigma_j\})\rrangle \equiv \bigotimes_{i=1}^{N/2}|\sigma_j\rrangle_{i}\otimes |\sigma_j\rrangle_{i+1}\;.
    \label{eq:q0q0}
\end{equation}
It follows that $\mathcal{W}_t$ is a sum of backward paths from each initial state $|\mathfrak{Q}_\mathrm{in}(\{\sigma_j\})\rrangle$ to each final $|\mathfrak{Q}_\mathrm{fin}(\{\sigma_j\})\rrangle \equiv \bigotimes_{i=1}^{N/2}|\tilde{\sigma}_j\rrangle_{i}\otimes |\sigma_j\rrangle_{i+1}$. 

We conclude this section highlighting the similarities and the differences of the IRP calculation with the computations for propagation of entanglement entropy~\cite{zhou2019,zhou2020}, defined as $S_q^\mathrm{ent} = -\ln [\mathrm{tr}(\rho_A^q)]$, where $\rho_A = \mathrm{tr}_{A_c}(\rho)$ is the reduced density matrix of the pure state $\rho=|\Psi\rangle\langle\Psi|$ and $A\cup A_c$ a bipartition of the system. 
As for the participation entropy, we focus on the annealed averages, determined by the purities, i.e., the reduced density matrix moments $\mathcal{P}_q \equiv \mathbb{E}_\mathrm{Haar}[\mathrm{tr}(\rho_A^q)]$. Denoting as $\mathfrak{I}\in \mathcal{S}_q$ the identity permutation $()$ and $\mathfrak{S}\in \mathcal{S}_q$ the cyclic permutation $(12\dots q)$, in the superoperator formalism we have 

\begin{equation}
  \mathcal{P}_q(t;A) = \llangle \mathfrak{Q}_\mathrm{in}(\{ \mathfrak{S}_j: j\in A, \mathfrak{I}_l: l\in A_c \})|\mathcal{W}_t^{(q)}|\psi_0^{\otimes N}\rrangle\;, 
\label{eq:enttensor}
\end{equation}
where $| \mathfrak{Q}_\mathrm{in}(\{ \mathfrak{S}_j: j\in A, \mathfrak{I}_l: l\in A_c \})\rrangle$ is a product of identity (cyclic) operators on the qudit $j$, for $j \in A_c$ ($j \in A$). 
The statistical mechanics model corresponding to the calculation of $\mathcal{P}_q(t;A)$ is exactly the same of the Eq.~\eqref{eq:Wtensor}, but with different boundary conditions: for the IPR we have free boundaries, cf. Eq.~\eqref{eq:q0q0}, while for the $\mathcal{P}_q(t;A)$ the boundary conditions are of the domain wall type. 
This fact leads to substantial differences between the Hilbert space delocalization and the entanglement propagation in the considered random circuits, which we will discuss in the next section focusing on the two-replica limit. 
We will further corroborate these findings with a numerical analysis for various replica number (R\'{e}nyi index) $q$.

\section{Two-replica computations}
\label{sec:randcircA}
In the two-replica limit, the parallelism between the inverse participation ratios and the reduced density matrix moments is more direct. Since $\mathcal{S}_2 = \{ \mathfrak{I},\mathfrak{S}\}$, Eq.~\eqref{eq:q0q0} translates to the sum over all domain wall boundary conditions~\eqref{eq:enttensor}. Therefore, the IPR, up to the overall constant, is given by  the sum of $\mathcal{P}_2$ over all possible choices of subsets $A$ of contiguous pairs of sites $\mathcal{Q}\equiv \{\{1,2\},\{3,4\},\dots,\{N-1,N\}\}$, cf. Eqs.~\eqref{eq:Wtensor} and Eq.~\eqref{eq:q0q0}, resulting in
%This results in the following formula for the circuit averaged IPR
\begin{equation}
    \overline{I}_2(t+1) = \frac{1}{(d^2+1)^{N/2}}\sum_{A\subset \mathcal{Q}} \mathcal{P}_2(t;A)\;,
    \label{eq:i2}
\end{equation}
where the shift to time $t+1$ comes from the simplification of one layer in Eq.~\eqref{eq:Wtensor}.
The time-evolution of the purities~\cite{zhou2019,zhou2020} can be split into two terms
\begin{equation}
    \mathcal{P}_2(t;A) = \frac{d^{2N_A} + d^{2(N/2-N_A)}}{d^N+1} + f_A(t)\;,
     \label{eq:e2}
\end{equation}
where the first time describes the long-time saturation value of the purity, the second term parametrizes the approach to the saturation value, and  the factors 2 two in the exponents arise 
%the pairs $\{i,i+1\}$ 
because the set $\mathcal{Q}$ contains $N/2$ pairs of neighboring lattice sites.
%Hilbert space dimensions in $\mathcal{Q}$. 
%The explicit solution for a single domain wall configuration, corresponding to a bipartition $A=\{1,2,\dots,2N_A\}$ and $A_c=\{2N_A+1,\dots,N\}$ with $N_A$ even, detailed in App.~\ref{app:B}
Combining the above Eqs.~\eqref{eq:i2} and~\eqref{eq:e2} leads to 
\begin{equation}
    \overline{I}_2(t+1) =  \frac{1}{(d^2+1)^{N/2}}\sum_{A\subset \mathcal{Q}}\left(\frac{d^{2N_A} + d^{2(N/2-N_A)}}{d^N+1} + f_A(t)\right) =  I^\mathrm{H}_2 + \frac{1}{(d^2+1)^{N/2}}\sum_{A\subset \mathcal{Q}}  f_A(t)\;,
\end{equation}
where we used that the first term in Eq.~\eqref{eq:e2} is independent of the partition~\footnote{This fact implies that 
\begin{equation}
   \sum_{A\subset \mathcal{Q}}{d^{2N_A} + d^{2(N/2-N_A)}}=  2\sum_{N_A=0}^{N/2} \binom{N/2}{N_A}d^{2N_A} = 2(d^2+1)^{N/2}\;.
   \nonumber
\end{equation}
}.
Converting to participation entropies, we have
\begin{equation}
    \tilde{S}^\mathrm{H}_2 - \tilde{S}_2(t) = 
    \ln\left[1 + \frac{d^N+1}{2(d^2+1)^{N/2}}\sum_{A\subset \mathcal{Q}}  f_A(t-1) \right]
    \label{eq:hatethis}
\end{equation}
%The first subleading contribution comes from the propagation of a single domain wall. From the large $d$ expansion, we get that 
We are interested in identifying the leading terms governing the decay of $\tilde{S}^\mathrm{H}_2 - \tilde{S}_2(t)$ at long times. Fixing the system size $N$ and considering the limit of $d \gg N$, the factor $(d^N+1)/[2 (d^2+1)^{N/2}]$ in Eq.~\eqref{eq:hatethis} simplifies to unity. We focus first on the single domain wall configurations, corresponding to bipartitions of the system into $A=\{1,2,\dots,2N_A\}$ and $A_c=\{2N_A+1,\dots,N\}$. The exact solution for the single domain wall case, detailed in App.~\ref{app:B}, shows that $f_A(t) \propto (2 K_d)^t$ for domains localized in the bulk of the system, where
\begin{equation}
      K_d = \frac{d}{d^2+1}.\;
     \label{eq:predd0}
\end{equation}
The time evolution of the purities $P_2(t;A)$ where $A$ corresponds to a configuration with $n_d$ domains is suppressed by the factor $(2K_d)^{n_d t}$, which vanishes at any $t>0$ in the large $d$ limit. Therefore, the leading contributions to the $ \tilde{S}^\mathrm{H}_2 - \tilde{S}_2(t)$ arise due to the single domain configurations. The number of the relevant single domain wall configurations scales proportionally to the system size $N$, translating Eq.~\eqref{eq:hatethis} to
\begin{equation}
     \tilde{S}^\mathrm{H}_2 - \tilde{S}_2(t)
     %\mapsto_{d\gg 1} 
     \stackrel{d \gg 1}{\longrightarrow}
     \alpha_d N (2K_d)^{t-1}
     %\;,\qquad K_d = \frac{d}{d^2+1}\;,
     \label{eq:predd}
\end{equation}
where the coefficient $\alpha_d
%\mapsto_{d\to\infty}=
 \stackrel{d \gg 1}{\longrightarrow}
1/2$ at sufficiently large system size $N$. 
While we have used the large on-site Hilbert space dimension limit to derive Eq.~\eqref{eq:predd}, we anticipate that this equation applies to any $d \geq 2$ with a properly chosen factor $\alpha_d$.

We are unable to analytically demonstrate the validity of Eq.~\eqref{eq:predd} at fixed finite value of $d$ and for $N \gg 1$. Nevertheless, we outline the underlying calculations which allow us to pin-point the leading factors in the time dependence of $\tilde{S}^\mathrm{H}_2 - \tilde{S}_2(t)$. First, we observe that at fixed $d$ and for $N \gg 1$, the contributions from the exponentially many configurations $A\subset \mathcal{Q}$ are exponentially in $N$ suppressed by the term $(d^N+1)/[2 (d^2+1)^{N/2}]$. Performing tensor network contractions, we find for a generic initial configuration that $f_A(t)=a_{A} (2K_d)^t$ in the long time limit. Interpreting the tensor network contraction as the wandering domain walls problem (cf. App.~\ref{app:B}), the factor $(2K_d)^t$ corresponds to the single domain wall configuration which dominates $f_A(t)$ at long times, while the factor $a_{A}$ depends on the processes required to reach the single domain wall configuration from the initial condition. The summation over all initial conditions yields an overall factor $(2K_d)^t$ which describes the behavior of $\tilde{S}^\mathrm{H}_2 - \tilde{S}_2(t)$ at any $d \geq 2$. We conjecture that the competition of the renormalizing factors $a_{A}$ with the exponential terms appearing in the summation over $A\subset \mathcal{Q}$ yields a factor proportional to the system size $N$ at any $d \geq 2$. This conjecture is equivalent to the validity of Eq.~\eqref{eq:predd} at any $d \geq 2$.

\begin{figure}[t!]
\centering
\includegraphics [width=1\textwidth]{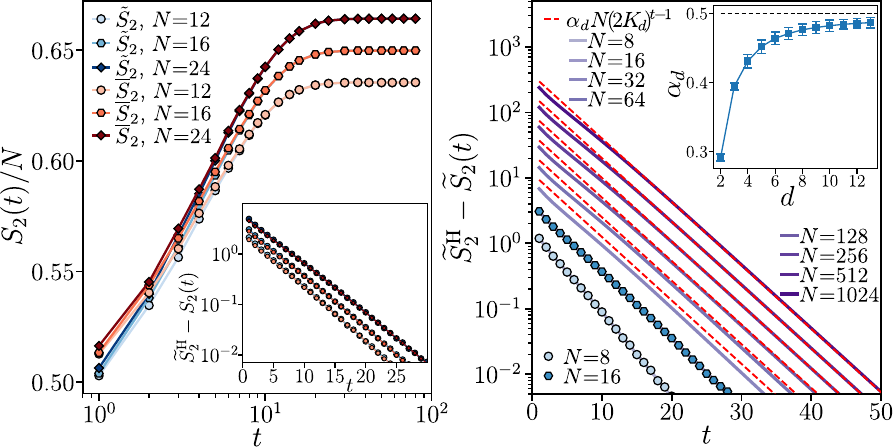}
\caption{
Participation entropy for the R\'{e}nyi index $q=2$. The left panel shows the quenched $\overline S_2$ (in red) and the annealed $\widetilde{S_2}$ (in blue) average of the participation entropy rescaled by the system size $N$ and plotted as a function of circuit depth $t$. The inset shows the exponential in time decay of the difference between the average participation entropy $\widetilde{S}^{\mathrm{H}}_2$ for random Haar states and the averages $\overline S_2, \widetilde{S}_2$ as a function of time. The right panel shows approach to $\widetilde{S}^{\mathrm{H}}_2$ of the annealed average $\widetilde{S_2}$ (blue symbols), the exact solution obtained with the tensor network approach (solid purple lines), compared with the  asymptotic formula $\widetilde{S}^{\mathrm{H}}_2 -\widetilde{S}_2(t) = \alpha_d {N} (2K_d)^{t-1}$ (red dashed line), where, for qubits, i.e. $d=2$, $\alpha_2\approx 0.291(5)$ and $K_2=\frac{2}{5}$. The inset shows the prefactor $\alpha_d$ as function of the on-site Hilbert space dimension $d$ the $\alpha_d $ extrapolated from tensor network simulations, while 
$\alpha \stackrel{d\to\infty}{\longrightarrow}\frac{1}{2}$
is shown by the black dashed line.
}\label{fig:num1}
\end{figure}

To corroborate %these results
our analytical considerations, we compare our predictions with the numerical simulations of the brick-wall random circuits varying the on-site Hilbert space dimension $d$.
A brief summary of our numerical implementations is detailed in App.~\ref{app:C}.
First, we numerically calculate the exact time evolution of the system's state $|\Psi \rangle$ up to the system size $N=24$. Our findings are reported in Fig.~\ref{fig:num1}(Left). We observe that already at the time scale $O(1)$ the difference between the annealed average $\overline{S}_2$ and the quenched average $\tilde{S}_2$ of the participation entropy is negligible.
Thus, the self-averaging properties of the IPR and participation entropies derived in the long-time limit in Sec.~\ref{sec:radomHaar} apply also for  the relatively shallow circuits allowing us to use the $\tilde{S}_2(t) = -\ln[\overline{I}_2(t)]$ as an accurate proxy for the circuit averaged participation entropy  $\overline{S}_2(t)$.

To test our analytical prediction  \eqref{eq:predd}, we focus on the difference between the stationary Haar value $\tilde{S}_2^\mathrm{H}$ and the annealed average $\tilde{S}_2(t)$. We implement Eq.~\eqref{eq:Wtensor} 
as a tensor network contraction, which allows us to reach system sizes $N\le 1024$ for any on-site Hilbert space dimension $d$. In Fig.~\ref{fig:num1}(Right) we observe that the difference $\tilde{S}_2^\mathrm{H}-\tilde{S}_2(t)$ decreases exponentially in time (circuit dept) $t$ proportionally to $(2K_d)^{t-1}$ or at slightly larger pace at the smallest considered system sizes $N=8,16$. 
The prefactor of the exponential decay increases monotonically with system size $N$. The growth of the prefactor becomes clearly linear beyond $N=64$. At $N > 64$ the behavior of  $\tilde{S}_2^\mathrm{H}- \tilde{S}_2(t)$  at longer times is captured with a good accuracy by Eq.~\ref{eq:predd} with $\alpha_2 = 0.291(5)$.

Performing tensor network calculations for the on-site Hilbert space dimension $d=3, \ldots, 13$ and for system sizes up to $N=256$, we verify that $\tilde{S}^\mathrm{H}_2 - \tilde{S}_2(t)$ decreases exponentially in time matching the prediction in Eq.~\eqref{eq:predd}, and reproducing the analytically found value of the constant $K_d$. We also find that 
the prefactor $\alpha_d$ increases with $d$, as shown in the inset in Fig.~\ref{fig:num1}(Right).
In particular, the numerical results indicate that $\alpha_d 
\stackrel{ d \to \infty }{\longrightarrow}1/2$, in accordance with the analytical prediction for the large on-site Hilbert space dimension limit.

The results discussed in this section demonstrate that the Hilbert space delocalization under the brick-wall random quantum circuits is an abrupt process. Indeed, for a given tolerance error $\varepsilon \ll 1$, the
participation entropy reaches $\tilde{S}_2(t) = \tilde{S}_2^\mathrm{H}-\varepsilon$ at time $t_\mathrm{HSD}\sim \ln(N)$ which scales logarithmically with the system size $N$.

\section{Numerical results for any replica}
\label{sec:randcircC}
For larger values of the R\'{e}nyi index $q\ge 3$, 
the analysis presented above cannot be trivially extended. 
First, the calculation of $\overline{I}_q$ requires all domain wall configurations $|\mathfrak{Q}_0(\{\sigma_j\})\rrangle$ with $\sigma_j\in \mathcal{S}_q$. Thus, the calculation of IPR involves a sum over many more initial configurations than just the configurations appearing in the computation of the purities $\mathcal{P}_q$ which  include only the identity permutation $\mathfrak{I}$ and the cyclic permutation $\mathfrak{S}$.
Moreover, $W^{(q)}_2$ involves contributions with negative signs and additional weights 
which affect the value of $K_d$, resulting in more complex analytical analysis.

\begin{figure}[t!]
\centering
\includegraphics [width=1\textwidth]{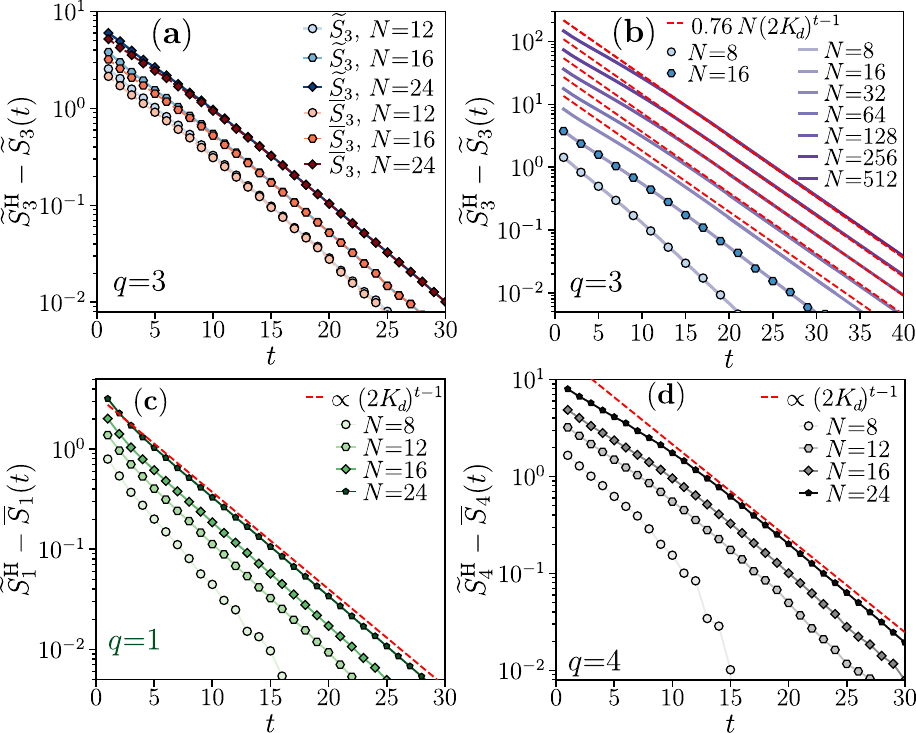}
\caption{
Participation entropy growth for various R\'{e}nyi indices $q$. Panel (a) shows the approach of the quenched $\overline S_3$ (in red) and annealed $\widetilde{S_3}$ (in blue) average of the participation entropy to the random Haar state value $\widetilde{S}^{\mathrm{H}}_3$ as a function of the circuit depth $t$. Panel (b) presents the approach to $\widetilde{S}^{\mathrm{H}}_3$ of the annealed average $\widetilde{S_3}$ (blue symbols) calculated with the exact numerical simulation compared with the tensor network contraction results (purple lines), and the fitted formula $\widetilde{S}^{\mathrm{H}}_3 -\widetilde{S}_3(t) \approx 0.76 N (2K_d)^{t-1}$ (red dashed line) consistent with Eq.~\eqref{eq:ddd}. Panels (c) and (d) show the approach of the circuit averaged participation entropy $\overline{S}_q$ to the random Haar state value $\widetilde{S}^{\mathrm{H}}_q$ respectively for $q=1$ and $q=4$.
}\label{fig:num2}
\end{figure}

Nonetheless, the main features of participation entropy growth under brick-wall quantum circuits discussed in Sec.~\ref{sec:randcircA} may be expected to hold for a generic replica number $q$. This leads us to the following conjecture 
\begin{equation}
    \tilde{S}_q^\mathrm{H}-\tilde{S}_q(t) \simeq  \alpha_{d,q} N \beta_{d,q}^t\;\qquad  \tilde{S}_q^\mathrm{H}-\tilde{S}_q(t)=\varepsilon\quad \text{at }t_\mathrm{HSD}\sim \ln(N)\;,
    \label{eq:ddd}
\end{equation}
where $\alpha_{d,q}$ is a coefficient such that $\alpha_{d,2} =\alpha_d$ in Eq.~\eqref{eq:predd}, and $\beta_{d,q}<1$ is a constant with $\beta_{d,2} = 2K_d$ and $\varepsilon \ll 1$ is a fixed tolerance.
The heuristic idea behind the above conjecture is that the limit $d\gg 1$ leads to positive weights in $W^{(q)}_2$, and up to corrections in $O(1/d)$, the main argument leading to Eq.~\eqref{eq:predd} should apply for Eq.~\eqref{eq:ddd}. 
We corroborate this hypothesis analyzing the dynamics for various values of $q$. The results are summarized in Fig.~\ref{fig:num2}.

For qubits ($d=2$), we first benchmark the self-averaging properties of the circuit, as shown in Fig.~\ref{fig:num2}(a) for $q=3$. The self-averaging holds similarly as for $q=2$, i.e. the annealed $\tilde{S}_3$ and the quenched average $\overline S_3$ of participation entropy rapidly approach each with the increase of the circuit's depth. We find analogous results for $q=1,4$ (data not shown), and expect that a similar phenomenology will arise for any R\'{e}nyi index $q$.
We therefore limit our discussion to the difference between the saturation value and the annealed average of the participation entropy, i.e., $\tilde{S}_q^\mathrm{H}-\tilde{S}_q(t)$. 

In panels (b,c,d) of Fig.~\ref{fig:num2} we demonstrate that 
the exponential decay of $\tilde{S}_q^\mathrm{H}-\tilde{S}_q(t)$ with the circuit depth $t$ is observed for all considered values of the R\'{e}nyi index $q$. For $d=2$ and $q=3$, see Fig.~\ref{fig:num2}(b), our tensor network implementation allows us to reach $N\le 512$. Surprisingly, we find that the exponential decay of $\tilde{S}_q^\mathrm{H}-\tilde{S}_q(t)$ is very accurately fitted with $\alpha_{2,3} N (2K_2)^t$, despite we could have expected a decay 
$\beta_{2,3}^t$ with a different coefficient $\beta_{2,3}$ fixed by the weights in $|\tilde{\sigma}\rrangle_{i,i+1}$, cf. Eq.~\eqref{eq:ddd} and Sec.~\ref{sec:randcirc}. 

For larger values of $q\geq 4$, as well as for the Von Neumann entropy limit $q=1$, we are limited to the relatively small system sizes $N\le 24$ accessible with the exact numerical simulation of the brick-wall random circuits. 
Time evolution of $\tilde{S}_q^\mathrm{H}-\tilde{S}_q(t)$ for $q=1,4$ approaches, with increasing system size $N$, the exponential decay $(2K_2)^t$. The separation of $\tilde{S}_q^\mathrm{H}-\tilde{S}_q(t)$ at a fixed time for different system sizes $N$ suggests that the prefactor in front of the exponential decay $(2K_2)^t$ may be scaling proportionally to the system size $N$. However, the range of available system sizes does not allow us to fully confirm this observation.

Finally, performing exact numerical calculations for $d=3$ (up to $N=14$) and $d=4$ (up to $N=10$) we verify that the decay of $\tilde{S}_q^\mathrm{H}-\tilde{S}_q(t)$ is accurately fitted with $(2K_d)^t$ for $q=1$ and $q=4$ (data not shown). This finding is analogous to the results for qubits $(d=2)$ and corroborates further the observation that the base $2K_d$ characterizing the exponential decay of $\tilde{S}_q^\mathrm{H}-\tilde{S}_q(t)$ is independent of the R\'{e}nyi index $q$.

Overall, the numerical results for the R\'{e}nyi index $q\neq 2$ presented in this section support our analytical analysis confirming that the approach of the participation entropy $\tilde{S}_q(t)$ to its long-time saturation value $\tilde{S}_q^\mathrm{H}$ occurs exponentially in time and with the prefactor that scales extensively with the number of qubits. Hence, also for $q \neq 2$ the approach, up to a fixed tolerance, of the participation entropy to its saturation value $\varepsilon$ occurs at time $t_\mathrm{HSD}\sim \ln(N)$ which scales logarithmically with the system size $N$.

\section{Discussion and conclusion}
\label{sec:conclusion}
We have investigated how the dynamics of random unitary circuits delocalize an initially localized state over a basis of the many-body Hilbert space. We combined analytical and numerical methods to calculate the time evolution of the participation entropies \eqref{eq:def} that characterize the spread of state $\rho = | \Psi \rangle \langle  \Psi |$ over a basis $\mathcal B$ of many qudit systems under dynamics of brick-wall random circuits. Our main finding is that the process of Hilbert space delocalization occurs abruptly so that the long-time saturation value of participation entropies is approached up to a fixed tolerance at times $t_\mathrm{HSD}\sim \ln(N)$ scaling logarithmically with the system size $N$. These results may appear surprising from the point of view of the close relations between the participation entropy and the entanglement entropy~\cite{mace2019multifractal, Sierant2022, detomasi, roy2022hilbertspace}. Indeed, entanglement entropy under local circuits grows linearly in time~\cite{nahum2017quantum}, saturating at times scaling proportionally to the system size $N$. While our analytical considerations show close parallels between entanglement entropy and participation entropy calculation, the boundary conditions of the relevant statistical mechanics model are different. This difference amounts to the uncovered contrasts between the time evolution of participation entropy and entanglement entropy. 
We focused on times short compared to the system size. A complementary path, considering random circuits in the limit of $t \to \infty$ and $N \to \infty$ and demonstrating the onset of universal features in distributions of the overlaps between the output states generalizing the Porter-Thomas distribution was considered in Ref.~\cite{andreaDluca}.

Random quantum circuits may be perceived as minimal models of local unitary dynamics. Hence, similarly to the entanglement entropy case~\cite{Kim13}, we expect that our results about Hilbert space delocalization extend to generic chaotic non-integrable many-body systems. Moreover, our results may provide a relevant reference point for understanding the Hilbert space delocalization in non-ergodic many-body systems, cf.~\cite{Torres-Herrera18, Creed23transportFS, Hopjan23survival}. The statistical invariance of the considered circuits under a product of arbitrary on-site unitary transformations shows that our results about the growth of participation entropies hold for any basis $\mathcal B_U$ obtained from the computational basis $\mathcal B$ by a product of independent on-site rotations.

The participation entropies considered here are not only directly available in numerical simulations of many-body systems but are also of experimental relevance. Indeed, recent progress in stochastic sampling of many-body wave functions in ultracold atomic and solid state experiments~\cite{Brydges19, Ebadi2021} allows, in principle, for their direct experimental evaluation. Similarly, the quantum processors realizing quantum circuits enable high frequency sampling of the output state~\cite{Arute2019, wu2021strong}, which is a basis of the cross-entropy benchmarking~\cite{Boixo2018}, and could enable a direct measuring of the participation entropies. Finally, the process of estimation of IPR, which, in general requires resources scaling exponentially with the system size $N$, may be simplified by use of appropriate quantum algorithms~\cite{YingjianLiu}.

This work has identified the leading terms relevant to the participation of entropy growth in random unitary dynamics. There are several interesting directions for further research. Analogously to the case of entanglement entropies~\cite{nahum2017quantum, entanglementgrowth}, one may ask a question about fluctuations of the participation entropies around the identified mean values. Properties of the dynamics of participation entropies under higher dimensional circuits are another unresolved problem. Moreover, stabilizer R\'enyi entropies~\cite{leone2022stabilizerrenyientropy,
haug2023efficient, oliviero2022measuringmagicon, gu2023little, tirrito2023quantifying}, which determine the amount of beyond-classical (non-Clifford) operations needed to perform a quantum task are related to participation entropies~\cite{turkeshi2023measuring}. The long long time limit of Stabilizer R\'enyi entropies under local unitary dynamics has been recently understood in~\cite{turkeshi2023pauli}. However, understanding the properties of growth of Stabilizer R\'enyi entropies is an exciting direction for further research, facilitated by the results of this work.

\authorcontributions{
X.T. commenced the project, performed analytical calculations and tensor network contraction computations. P.S. performed exact numerical simulations. Both authors contributed to discussion of the results and writing of the manuscript.}

\funding{X.T. acknowledges DFG under Germany's Excellence Strategy – Cluster of Excellence Matter and Light for Quantum Computing (ML4Q) EXC 2004/1 – 390534769, and DFG Collaborative Research Center (CRC) 183 Project No. 277101999 - project B01. PS acknowledges support from European Research Council AdG NOQIA; MCIN/AEI (PGC2018-0910.13039/501100011033, CEX2019-000910-S/10.13039/501100011033, Plan National FIDEUA PID2019-106901GB-I00, Plan National STAMEENA PID2022-139099NB, I00,project funded by MCIN/AEI/10.13039/501100011033 and by the “European Union NextGenerationEU/PRTR" (PRTR-C17.I1), FPI); QUANTERA MAQS PCI2019-111828-2); QUANTERA DYNAMITE PCI2022-132919, QuantERA II Programme co-funded by European Union’s Horizon 2020 program under Grant Agreement No 101017733); Ministry for Digital Transformation and of Civil Service of the Spanish Government through the QUANTUM ENIA project call - Quantum Spain project, and by the European Union through the Recovery, Transformation and Resilience Plan - NextGenerationEU within the framework of the Digital Spain 2026 Agenda; Fundació Cellex; Fundació Mir-Puig; Generalitat de Catalunya (European Social Fund FEDER and CERCA program, AGAUR Grant No. 2021 SGR 01452, QuantumCAT \ U16-011424, co-funded by ERDF Operational Program of Catalonia 2014-2020); the computing resources at Urederra and technical support provided by NASERTIC (RES-FI-2024-1-0043); Funded by the European Union. Views and opinions expressed are however those of the author(s) only and do not necessarily reflect those of the European Union, European Commission, European Climate, Infrastructure and Environment Executive Agency (CINEA), or any other granting authority. Neither the European Union nor any granting authority can be held responsible for them (EU Quantum Flagship PASQuanS2.1, 101113690, EU Horizon 2020 FET-OPEN OPTOlogic, Grant No 899794), EU Horizon Europe Program (This project has received funding from the European Union’s Horizon Europe research and innovation program under grant agreement No 101080086 NeQSTGrant Agreement 101080086 — NeQST); ICFO Internal “QuantumGaudi” project; European Union’s Horizon 2020 program under the Marie Sklodowska-Curie grant agreement No 847648; “La Caixa” Junior Leaders fellowships, La Caixa” Foundation (ID 100010434): CF/BQ/PR23/11980043.}

\dataavailability{The code for the numerical simulations and the data are available at [ref].} 

\acknowledgments{We thank A. De Luca and L. Piroli for enlightning discussions, and collaborations on related topics. 
}

\conflictsofinterest{The authors declare no conflicts of interest.} 

%%%%%%%%%%%%%%%%%%%%%%%%%%%%%%%%%%%%%%%%%%
%% Optional

%%%%%%%%%%%%%%%%%%%%%%%%%%%%%%%%%%%%%%%%%%
%% Optional
\appendixtitles{no} % Leave argument "no" if all appendix headings stay EMPTY (then no dot is printed after "Appendix A"). If the appendix sections contain a heading then change the argument to "yes".
\appendixstart
\appendix
\section[\appendixname~\thesection]{Schur-Weyl duality and integration over Haar unitaries}
\label{app:A}
In this Appendix we briefly review the average over $q$-tensor product of Haar distributed unitary $U\in \mathcal{U}(d^N)$. Here, $\mathcal{U}(d^N)$ is the unitary group acting on $N$ qudits with local Hilbert space dimension $d$. At the heart of our discussion is the averaged channel 
\begin{equation}
   \Phi^{(q)}_{\mathrm{Haar}}(\bullet) = \mathbb{E}_{\mathrm{Haar}}\left[ U^{\otimes q} (\bullet) (U^\dagger)^{\otimes q}  \right] \equiv \int\; d\mu(U) U^{\otimes q} (\bullet)(U^\dagger)^{\otimes q}\;,
\end{equation}
where $\mu(U)$ is the Haar measure over $\mathcal{U}(d^N)$, whose defining feature is the invariance under left group multiplication: $\mu(U) = \mu(VU)$ for any $V\in \mathcal{U}(d^N)$. 
It is noteworthy that $\Phi^{(q)}_{\mathrm{Haar}}(\bullet)$ commutes with all unitaries $V^{\otimes q}$ with $V\in \mathcal{U}(d^N)$, a fact following from 
\begin{equation}
\begin{split}
    V^{\otimes q} \Phi^{(q)}_{\mathrm{Haar}}(\bullet) (V^\dagger)^{\otimes q} &= \int_{\mathrm{Haar}} d\mu(U) V^{\otimes q} U^{\otimes q}   (\bullet)  (U^\dagger)^{\otimes q} (V^\dagger)^{\otimes q} \\
    &= \int_{\mathrm{Haar}} d\mu(\tilde{U})  \tilde{U}^{\otimes q} (\bullet)  (\tilde{U}^\dagger)^{\otimes q} = \Phi^{(q)}_{\mathrm{Haar}}(\bullet)\;.
\end{split}
\end{equation}

By Schur-Weyl duality, it then can be expressed in terms of permutation operators
\begin{equation}
    \Phi^{(q)}_{\mathrm{Haar}}(\bullet) = \sum_{\sigma,\tau \in \mathcal{S}_q} \mathrm{Wg}(d^N;\sigma \tau^{-1}) \mathrm{tr}\left(T_\sigma (\bullet)\right) T_\tau \;,
    \label{eq:weingarten}
\end{equation}
where $\mathcal{S}_q$ is the permutation group over $q$ elements, and each permutation $\sigma \in \mathcal{S}_k$ is represented as a replica operator $T_\sigma |n_1,\dots,n_q\rangle = |n_{\sigma(1)},\dots, n_{\sigma(q)}\rangle$, for any basis $\{n_j=0,\dots,d^N-1\}$ over the many-body Hilbert space $\mathcal{H}_N=  \mathbb{C}^{d N}$. Without loss of generality, guaranteed by the Haar invariance of $\Phi^{(q)}_{\mathrm{Haar}}(\bullet)$, we can focus on the computational basis. Here, each $n_j$ is represented by the $d$-modular string $|n_j\rangle= |b_j^1,\dots,b_j^N\rangle$, with $b_j^k=0,\dots,d-1$. It is then easy to see that $T_\sigma=t_\sigma^{\otimes N}$ is reducible to permutation operators acting on each qudit digit, namely $t_\sigma^{(j)} |b^1_j,b^2_j,\dots,b^q_j\rangle =|b^{\sigma(1)}_j,b^{\sigma(2)}_j,\dots,b^{\sigma(q)}_j\rangle$. (We will occasionally use the superscript $(j)$ to clarify upon which qudit the action is on, stressing that the operator representation is the same.)

The expression in Eq.~\eqref{eq:weingarten} was originally determined in [refs], and $\mathrm{Wg}(d^N;\sigma \tau^{-1})$ are known as Weingarten coefficients. They are implicitly determined from $\Phi_{\mathrm{Haar}}^{(q)}(T_\tau) = T_\tau$, leading to linear equation
\begin{equation}
    \delta_{\sigma,\tau} = \sum_{\omega\in \mathcal{S}_q} \mathrm{Wg}(d^N;\sigma \omega^{-1}) d^{N \#(\omega \tau)}\;,
\end{equation}
where we used $\mathrm{tr}(t_\omega t_\tau) = d^{\#(\omega \tau)}$, and $\#(\sigma)$ is the number of cycles of $\sigma$. 
Passing to the superoperator formalism, we have $t_\tau^{(j)} \mapsto |\tau\rrangle_j$. Furthermore, since $U^{\otimes q}(\bullet) (U^\dagger)^{\otimes q}\mapsto (U\otimes U^*)^{\otimes q} (\bullet)$, it follows that $\Phi^{(q)}_{\mathrm{Haar}}(\bullet)\mapsto W^{(q)}_N$, cf. Eq.~\eqref{eq:WWW}. There, the decomposition in terms of the dual operators is recast using 
\begin{equation}
    \llangle \tilde{\sigma} |_{1,\dots,N} = \sum_{\tau \in \mathcal{S}_q} \mathrm{Wg}(d^N;\sigma \tau^{-1}) \llangle \tau|_1\otimes \cdots \otimes \llangle \tau|_N\;.
\end{equation}

\begin{figure}[t!]
\centering
\includegraphics [width=0.7\textwidth]{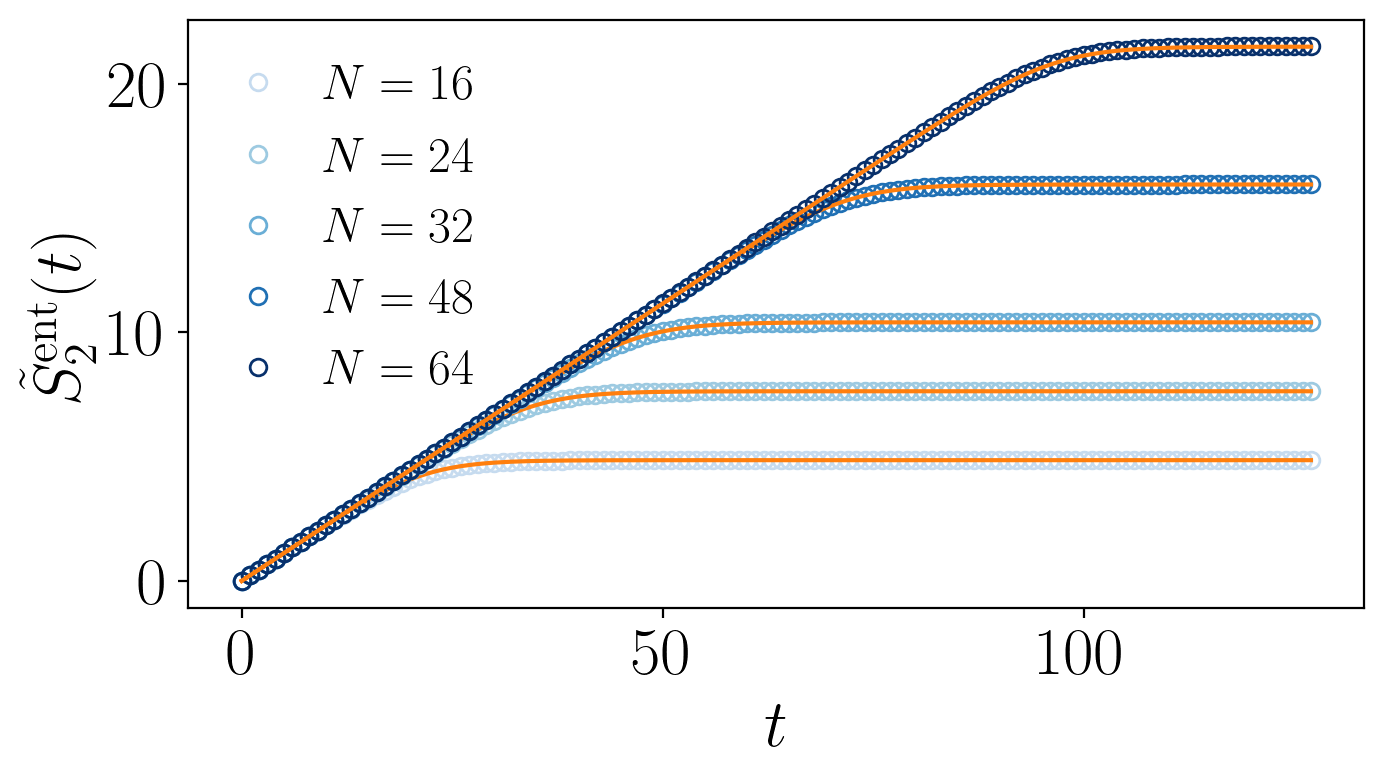}
\caption{
Annealed averaged R\'enyi-2 entanglement entropy for $N_A = N/2$. Marker are numerical simulations obtained with tensor networks and orange lines are the exact expression Eq.~\eqref{eq:exP}, in quantitative agreement.
}\label{fig:entropia}
\end{figure}

\section[\appendixname~\thesection]{Entanglement growth for finite circuits}
\label{app:B}
We detail the computation of the annealed averaged entanglement entropy. 
We consider the bipartition $A=\{1,\dots,N_A\}$ and $A_c = \{N_A+1,\dots,N\}$. In this case, the average purity of the reduced density matrix evolve as 
\begin{equation}
    \mathcal{P}_2(t) = \llangle \mathfrak{S}^{\otimes N_A}\mathfrak{I}^{\otimes (N-N_A)}|\mathcal{W}_t|\psi_0,\dots,\psi_0\rrangle\;.
\end{equation}
Recalling the structure in Eq.~\eqref{eq:Wtensor} and that  
\begin{equation}
    W_2^{(2)} = \begin{pmatrix}
        1 & K_d & K_d & 0 \\
        0 & 0 & 0 & 0 \\
        0 & 0 & 0 & 0 \\
        0 & K_d & K_d & 1 \\
    \end{pmatrix}\;,
\end{equation}
it follows that $\mathcal{P}_2(t)$ is mapped to a  random walk (of the entanglement cut), with absorbing boundaries at zero and $N$. Let us define the probability $u_{z,t}$ that the random walk is absorbed exactly at time $t$ when starting from the position $z=N_A$. It follows that
\begin{equation}
    u_{z,t+1} = \frac{1}{2} u_{z-1,t} + \frac{1}{2} u_{z+1,t}\;,\qquad u_{z,0} = \delta_{z,0} + \delta_{z,N}\;,\qquad  u_{0,t} = u_{N,t} = 1\;.
\end{equation}
Solving this stochastic process we have
\begin{equation}
    u_{z,t} = \sum _{\nu =0}^{N/2-1} \frac{2}{N}\sin \left(\frac{\pi  (2 \nu +1)}{N}\right) \cos ^{t-1}\left(\frac{\pi  (2 \nu +1)}{N}\right) \sin \left(\frac{\pi  (2 \nu +1) z}{N}\right)\;.
\end{equation}
Since each drift has a energy cost of $2K$, we have the final result 
\begin{equation}
    \mathcal{P}_2(t) = (2K_d)^t \sum_{s=t+1}^\infty u_{N_A,s} + \sum_{s=0}^t (2K_d)^s u_{N_A,s}\;. 
    \label{eq:exP}
\end{equation}
In particular, for $N\to\infty$ we recast the findings in Ref.~\cite{zhou2020}, and for sufficiently short time $t\lesssim N/2$, we find that $S^\mathrm{ent}_2 =  v_d t$, with the entanglement velocity $v_d = -\ln(2K)$. 

\section[\appendixname~\thesection]{Details of the numerical implementation}
\label{app:C}
This Appendix discusses briefly the numerical methods employed in this manuscript.

\paragraph{Exact computational methods} 
For generic values of $q$ and qubit ($d=2$) circuits, we evolve the exact representation of the state on the full Hilbert space. Each gate action translates to a sparse matrix multiplication, resulting for $N$ qubits and $M$ gates to $O(2^N)$ memory and $O(M 2^N)$ runtime requirements. 
The knowledge of the state at each time step (circuit depth), allows us to compute $I_q$ for any value of $q$. We store the values of $I_q(t)$ and $S_q(t)$ for each realization of the circuit, and average over $\mathcal{N}_H=10^4$ realizations. This allows us to obtain both annealed and quench average of the participation entropies.

\paragraph{Tensor network simulations}
For integer $q$, we can represent the average circuit in the tensor network formalism. As discussed in  Sec.~\ref{sec:randcirc}, the local degree of freedom is a "spin" of dimension $d_\mathrm{eff}=q!$. Similarly, the $W_2^{(q)}$ transfer matrix is a $d_\mathrm{eff}\times d_\mathrm{eff}$ gate. 
We store the initial state $|\mathfrak{Q}_\mathrm{in}\rrangle$, cf.  Sec.~\ref{sec:randcircA}, in a matrix product state (MPS) ~\cite{Schollw_ck_2011}
\begin{equation}
    |\mathfrak{Q}_\mathrm{in}\rrangle = \sum_{\{s_j=0,\dots,d_\mathrm{eff}-1\}} A_{1,\alpha_1}^{s_1}A_{\alpha_1,\alpha_2}^{s_2}\cdots A_{\alpha_{N-1},1}^{s_N}|s_1,\dots,s_N\rrangle\;.
\end{equation}
Each layer in $\mathcal{W}_t$ is an MPO, which we apply to the state, and compress with tolerance $\varepsilon=10^{-15}$. The final contraction is with the state $|\mathfrak{Q}_\mathrm{fin}\rrangle$. Our implementation is based on the opensource library \textsc{ITensor}~\cite{itensor}.

\begin{adjustwidth}{-\extralength}{0cm}

\reftitle{References}


\bibliography{newbib}

\PublishersNote{}
\end{adjustwidth}
\end{document}